\documentstyle[aps,prl%
,preprint%
]{revtex} 
\preprint{Appears in {\it Phys. Rev. Lett.}}
\begin{document} 
\draft
\title{DIRECT HOPF BIFURCATION IN PARAMETRIC RESONANCE OF 
   HYBRIDIZED WAVES}
\author{Franz-Josef Elmer}
\address{Institut f\"ur Physik, Universit\"at 
   Basel, CH-4056 Basel, Switzerland}
\date{January 29, 1996}
\maketitle
\begin{abstract}

We study parametric resonance of interacting waves having the same
wave vector and frequency. In addition to the well-known
period-doubling instability we show that under certain conditions the
instability is caused by a Hopf bifurcation leading to quasiperiodic
traveling waves. It occurs, for example, if the group velocities of
both waves have different signs and the damping is weak. The dynamics
above the threshold is briefly discussed. Examples concerning
ferromagnetic spin waves and surface waves of ferro fluids are
discussed.

\end{abstract}

\narrowtext

Parametric resonance is an instability phenomenon that is responsible
for the excitation of modes of a weakly damped system due to periodic
modulation of parameters which determine the frequencies of the modes
\cite{cro.93}. The prototypical example is the vertically driven
pendulum where the modulated parameter is the acceleration of gravity
\cite{lan.60}. In spatially extended systems these modes are waves. 
Experimentally parametric resonance is well-known for spin waves
(Suhl instabilities) \cite{suh.57,dam.63,lvo.94} and surface waves of
fluids (Faraday instability) \cite{far.31,mil.84,mil.91}.

Let us briefly summarize the well-known behavior of the one-wave case
where only one wave type is involved. The ground state (e.g., the
flat surface of a fluid) becomes unstable if the driving (i.e.,
modulation) amplitude exceeds some threshold which is proportional to
the damping constant. The instability is caused by the excitation of
waves fulfilling the first-order parametric resonance condition
\begin{equation}
  \Omega({\bf k})=\frac{\omega}{2},
  \label{res.con}
\end{equation}
where $\Omega({\bf k})$ is the frequency of the wave with wave number
${\bf k}$ (i.e., the dispersion relation) and $\omega$ is the
modulation frequency. At the instability threshold a period-doubling
bifurcation occurs because the wave solution has a period which is
two times the period of the modulation. That is, in the stroboscopic
map (i.e., the map which gives the state at time $t=2(m+1)\pi/\omega$
provided the state at $t=2m\pi/\omega$ is known) a period-two orbit
bifurcates.

How does this one-wave picture change if two types of waves with
different dispersion relations $\Omega_1$ and $\Omega_2$ are
involved? There is in principle no change as long as the manifolds
defined by $\Omega_j({\bf k})=\omega/2$ do not intersect. At such
intersection lines two different waves with the {\em same\/} wave
vector fulfill the parametric resonance condition. Interaction leads
to the well-known effect of {\em hybridization\/} where the pure
waves are replaced by mixtures of them. Although hybridization avoids
the intersection of the above-mentioned manifolds they still come
very close together with a minimum distance proportional to the
strength of interaction. Below we will see that under certain very
general conditions a direct Hopf bifurcation occurs. It leads to
quasiperiodic motion right at the onset of parametric resonance. 

The wave lengths of waves which are excited due to parametric
resonance are usually large compared to the atomic scales. Their
dynamics are described by macroscopic field equations like the
Navier-Stokes equations for fluids or the Landau-Lifshitz equations
for magnets.  Formally the fields can be put together into a single
vector field ${\bf u}({\bf r},t)$, called order parameter. In the
limit of weak damping universal equations of motion so-called
amplitude equations can be derived \cite{cro.93}. They describe the
spatio-temporal evolution of the amplitudes of waves fulfilling the
parametric resonance condition. The amplitude equations are derived
from a multiple-scale perturbation theory where the damping constant
is the smallness parameter. For the field ${\bf u}$ we make the
ansatz:
\begin{eqnarray}
  {\bf u}({\bf r},t)&=&(A_+{\bf u}_++A_-{\bf u}_-)e^{i(kx-\omega 
   t/2)}+\nonumber\\&&+(B_+{\bf v}_++B_-{\bf v}_-)e^{i(kx+\omega 
   t/2)}+{\rm c.c.},
  \label{u}
\end{eqnarray}
where $A_\pm$ and $B_\pm$ are the slowly varying amplitudes of the
right and left waves, respectively. The corresponding wave solutions
are ${\bf u}_\pm$ and ${\bf v}_\pm$. In general they depend on $y$
and $z$. The index $\pm$ corresponds to the different types of waves.
The cross point of their dispersion relations $\Omega_\pm(k)$ are
denoted by $k_0$ and $\omega_0$, i.e., $\Omega_\pm(k_0)=\omega_0$. It
is assumed that the modulation amplitude $h$, the detuning
$\omega/2-\Omega(k)$, the amplitudes of the waves, and the coupling
strength between different types of waves are of first order in the
damping constant. In the one-wave case the amplitude equations for
$A_\pm$ and $B_\pm$ are well-known \cite{cro.93,dou.89,rie.90}.
Coupling between $+$ and $-$ waves lead to additional linear cross
terms. Because of the hamiltonian nature of these coupling ${\bf
u}_\pm$ and ${\bf v}_\pm$ can always be chosen in such a way that the
linear part of the amplitude equations reads:
\begin{mathletters}\label{amp.eq}
\begin{eqnarray}
   \dot A_\pm&=&-\Gamma_\pm A_\pm+ih\alpha_\pm B_\pm
      +i\epsilon e^{\pm i\chi}A_\mp,\\
   \dot B_\pm&=&-\Gamma^*_\pm B_\pm-ih\alpha_\pm A_\pm
      +i\epsilon e^{\mp i\chi}B_\mp,
\end{eqnarray}
\end{mathletters}
with
\begin{equation}
 \Gamma_\pm =\gamma_\pm+i(\Omega_\pm-\omega/2),\ \ \gamma_\pm>0,
 \label{Gamma}
\end{equation}
where $\gamma_\pm$, $\Omega_\pm$, and $\alpha_\pm$ are real numbers
denoting for each wave type the damping constant, the dispersion
relation, and the strength of modulation, respectively, $\epsilon$
and $\chi$ are the strength and the phase of the coupling,
respectively. These parameters are in general functions of the wave
number $k$. The amplitude and frequency of modulation are denoted by
$h$ and $\omega$, respectively.

The ground state $A_\pm=B_\pm=0$ becomes unstable if a solution
$A_\pm=a_\pm\exp(\lambda t)$, $B_\pm=b_\pm\exp(\lambda t)$ exists
with ${\rm Re}\,\lambda>0$. For increasing $h$, $\lambda$ will
eventually cross the imaginary axis defining a bifurcation point.
Since the coefficients of (\ref{amp.eq}) depend on $k$ these points
become functions of $k$ called {\em neutral curves\/}. There are two
different types of neutral curves possible: (i) Neutral curves
$h_S(k)$ where $\lambda=0$ (soft-mode or stationary instability) and
(ii) neutral curves $h_H(k)$ where ${\rm Im}\,\lambda\equiv
\omega_H\neq 0$ (hard-mode or oscillatory instability). In the
stroboscopic map they correspond to period-doubling bifurcation and
Hopf bifurcation, respectively. The neutral curves $h_S$ and $h_H$
are the real solutions of
\begin{equation}
  Eh^4-Fh^2+G=0
  \label{thr.eq}
\end{equation}
with $E$, $F$, and $G$ given by
\begin{mathletters}
\label{EFG.S}
\begin{eqnarray}
 E_S&=&(\alpha_+\alpha_-)^2,\\
 F_S&=&|\alpha_+\Gamma_-|^2+|\alpha_-\Gamma_+|^2-2
   \alpha_+\alpha_-\epsilon^2\cos 2\chi,\\
 G_S&=&|\Gamma_+\Gamma_-+\epsilon^2|^2,
\end{eqnarray}
\end{mathletters}
and
\begin{mathletters}
\label{EFG.H}
\begin{eqnarray}
 E_H&=&\gamma_+\gamma_-(\alpha_+^2-\alpha_-^2)^2,\\
 F_H&=&2\gamma_+\gamma_-[(\alpha_+^2-\alpha_-^2)
   (|\Gamma_+|^2-|\Gamma_-|^2)+\nonumber\\&&+2(\gamma_++\gamma_-)
   (\gamma_-\alpha_+^2+\gamma_+\alpha_-^2)]+\nonumber\\&&
   +(\gamma_++\gamma_-)^2(\alpha_+^2+\alpha_-^2+2\alpha_+\alpha_-\cos
   2\chi)\epsilon^2,\\
 G_H&=&\gamma_+\gamma_-[(|\Gamma_+|^2-|\Gamma_-|^2)^2+\nonumber\\&&
   +4(\gamma_++\gamma_-)(\gamma_-|\Gamma_+|^2+\gamma_+|\Gamma_-|^2)]
   \nonumber\\&&+|(\gamma_++\gamma_-)(\Gamma_++\Gamma_-)\epsilon|^2,
\end{eqnarray}
\end{mathletters}
respectively. The solution $h_H$ is relevant only if the Hopf
frequency $\omega_H$ given by
\begin{equation}
  \omega_H^2=\epsilon^2+\frac{\gamma_-|\Gamma_+|^2+\gamma_+
   |\Gamma_-|^2-(\gamma_-\alpha_+^2+\gamma_+\alpha_-^2)h_H^2}
   {\gamma_++\gamma_-}
  \label{w.h}
\end{equation}
is real.

In the noninteracting case (i.e., $\epsilon=0$) the characteristic
polynomial (\ref{thr.eq}) factorizes into two second-order
polynomials and we are back to the one-wave case. The instability is
always caused by a real $\lambda$ crossing the imaginary axes and
consequently $h_H$ does not exist. The neutral curve reads
$h_S=|\Gamma_\pm|/\alpha_\pm$. The instability threshold $h_c$ (i.e.,
the absolute minimum) is given by
$h_c=h_\pm\equiv\gamma_\pm/\alpha_\pm$.

The two-wave case (i.e., $\epsilon\not= 0$) is qualitatively much
richer. Outside the hybridization area the one-wave picture is still
valid.  We assume that the threshold is given by the $+$ wave, i.e.,
$h_c=h_+$.  How does the type and the threshold of the instability
change if we sweep through a hybridization area? First it can be
rigorously proved that the threshold is always higher than $h_+$.
This can be easily understood because the coupling of a
parametrically excited wave with another wave opens an additional
dissipation channel. Thus the driving amplitude has to be increased
in order to excite a wave parametrically. In the hybridization region
where the coupling is strongest this effect is most pronounced. In
the limit of strong damping or weak coupling this effect is of the
order $\epsilon^2$.  In this limit direct Hopf bifurcations are not
possible.

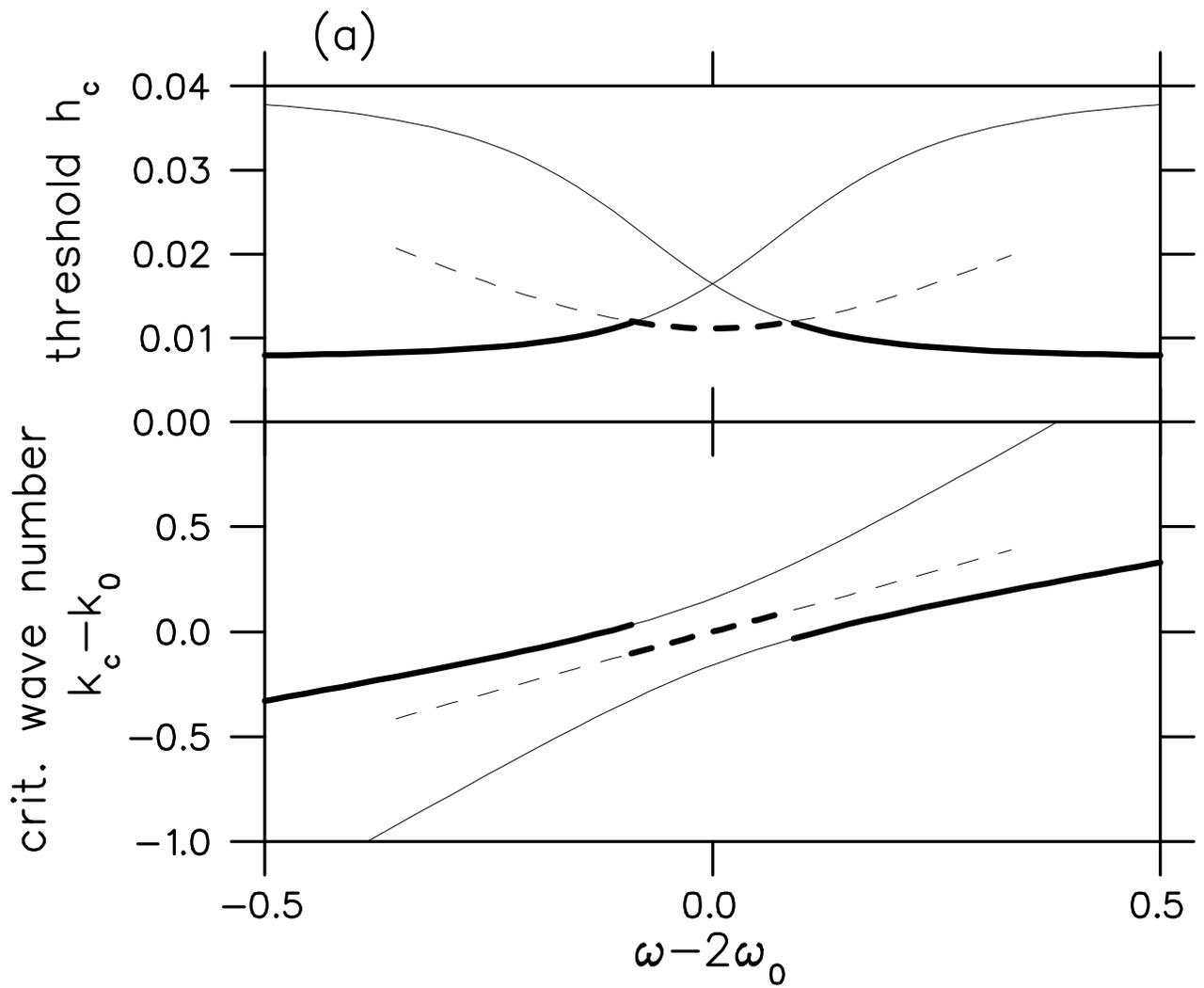
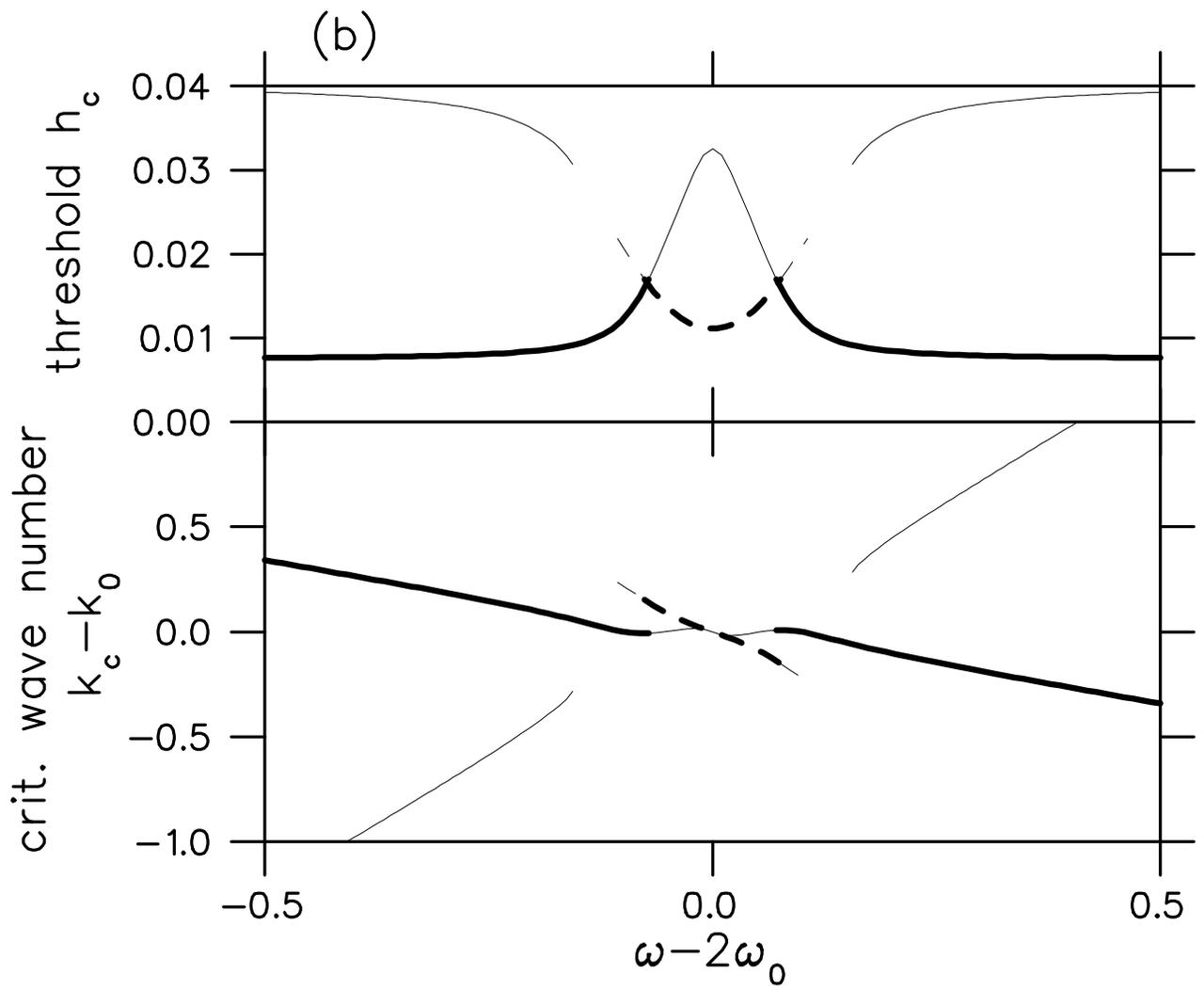
\begin{figure}
\caption[Threshold and kc]{\protect\label{f.qc}
Thresholds and critical wave numbers as functions of the modulation
frequency near a hybridization region. The parameters are
$\epsilon=0.06$, $\chi=1.5$, $\alpha_+=4$, $\alpha_-=0.5$,
$\gamma_+=0.03$, $\gamma_-=0.02$, $\Omega_-=\omega_0+0.2(k-k_0)$, and
(a) $\Omega_+=\omega_0+0.7(k-k_0)$, (b)
$\Omega_+=\omega_0-0.7(k-k_0)$.  Solid (dashed) lines indicate
relative minima of $h_S$ ($h_H$). The absolute minimum is denoted by
thick lines. 
}
\end{figure}

For the other limit (i.e., strong coupling or weak damping) we have
to distinguish between two cases depending on whether the slopes of
the dispersion relations (i.e., the group velocities $c_\pm\equiv
d\Omega_\pm/dk$) have the same sign or not. When both signs are
identical the neutral curve $h_S$ has always two relative minima
located near the parametric resonance condition. The reason for that
is that the
hybridized dispersion relations are monotonic functions of the wave
number. Sweeping $\omega$ through the hybridization region one
minimum increases whereas the other one decreases [see
Fig.~\ref{f.qc}(a)]. There will be a point in the middle near
$\omega=2\omega_0$ where both minima are of equal height. This is a
co-dimension two bifurcation point where a competition between two
period-doubling bifurcations for two different values of $k$ takes
place. Assuming no $k$-dependence of the group velocities $c_\pm$ and
the parameters $\gamma_\pm$, $\alpha_\pm$, $\epsilon$, and $\chi$ we
get approximatively for the threshold in the co-dimension two point
\begin{equation}
  h_{Sc}(2\omega_0)=\frac{|c_-\gamma_++c_+\gamma_-|}{\sqrt{c_-^2
   \alpha_+^2+c_+^2\alpha_-^2-2c_+c_-\alpha_+\alpha_-\cos 2\chi}}.
  \label{hc.2}
\end{equation}
The error is of quadratic order in the damping constants. Note that
the threshold enhancement $h_{Sc}(2\omega_0)-h_+$ can be of the same
order as $h_+$ [see Fig.~\ref{f.qc}(a)].

If the group velocities have different signs the hybridized
dispersion relations are no longer monotonic. Thus an interval around
$\omega=2\omega_0$ exists where the parametric resonance condition
can not be fulfilled. Therefore the threshold $h_S$ is much larger
than in the case of equal signs (see Fig.~\ref{f.qc}) because it can
not be of the order of the damping constants like (\ref{hc.2}). 

In the limit of strong coupling or weak damping the instability due
to a period-doubling bifurcation can be beaten by an oscillatory 
instability (see Fig.~\ref{f.qc}). Its neutral curve $h_H$ is very
well approximated by 
\begin{equation}
  h_H=\frac{|\Gamma_++\Gamma_-|}
   {\sqrt{\alpha_+^2+\alpha_-^2+2\alpha_+\alpha_-\cos 2\chi}}.
  \label{hH}
\end{equation}
The Hopf frequency $\omega_H$ is in leading order given by
$\epsilon$ which is half of the hybridization gap at $k=k_0$. The
threshold 
\begin{equation}
  h_{Hc}(\omega)=\frac{\gamma_++\gamma_-}
   {\sqrt{\alpha_+^2+\alpha_-^2+2\alpha_+\alpha_-\cos 2\chi}}
  \label{h.Hc}
\end{equation}
occurs at
\begin{equation}
  \Omega_+(k_{Hc})+\Omega_-(k_{Hc})=\omega.
  \label{k.Hc}
\end{equation}
Eq.~(\ref{h.Hc}) is a good approximation only near
$\omega=2\omega_0$.  Going away from this point $h_{Hc}(\omega)$
increases [see Fig.~\ref{f.qc}(b)]. It eventually disappears because
$h_H$ does not exists far away from the hybridization area.

What are the conditions for $h_{Hc}<h_{Sc}$? Because of too many
parameters we can not give a complete answer. We will consider only
two important case where an easy answer is possible. In the first
case the group velocities $c_+$ and $c_-$ have different signs and 
$h_{Hc}$ is almost always less than $h_{Sc}$ because $h_{Hc}\sim{\cal
O}(\gamma)$ and $h_{Sc}\sim{\cal O}(\gamma^0)$. For equal signs we
get an easy conditions if only one wave is driven parametrically
(i.e., $\alpha_-=0$). Here (\ref{hc.2}) and (\ref{h.Hc}) simplify to 
$h_{Sc}=(\gamma_++\gamma_-|c_+/c_-|)/\alpha_+$ and $h_{Hc}=(\gamma_+
+\gamma_-)/\alpha_+$, respectively. Thus the Hopf bifurcation occurs
first if $|c_-|<|c_+|$.

In a preliminary study we have also investigated the bifurcating 
solutions by including usual third-order terms in the amplitude
equations \cite{cro.93,dou.89,rie.90}. The bifurcation at $h_H$ is
either supercritical or weakly subcritical. The bifurcating solution
is similar to the drift solution which occurs in a secondary
instability in the one-wave case \cite{cro.93,dou.89}. For larger
values of $h$ this solution becomes oscillatory unstable. The
corresponding Hopf bifurcation leading to a torus solution is either
supercritical or weakly subcritical. Note that for the order
parameter ${\bf u}$ this solution contains three different
frequencies. Further increase of $h$ leads via a period-doubling
sequence to chaos. 

In which physical systems can we expect direct Hopf bifurcation due
to parametric resonance? First we distinguish two broad classes of 
physical systems where hybridization of waves on macroscopic scales
takes place. The first class is characterized by at least two
different types of waves caused by different physical mechanisms,
e.g., spin waves and sound waves in ferromagnets. The phonon-magnon
system has already been studied experimentally in ferromagnets by
parallel-pumping \cite{tur.60}. Parallel pumping means an additional
high-frequency magnetic field parallel to the static one. In
parallel-pumping only the spin waves are driven. Unfortunately the
usual period-doubling bifurcation comes first because the sound
velocity is larger than the magnon group velocity. But the threshold
enhancement has been observed \cite{tur.60}. We predict a
direct Hopf bifurcation if only sound waves are excited
parametrically.

The second class of systems with hybridization is obtained by
squeezing a wave with an anisotropic and/or nonmonotonic dispersion
relation into a geometry with small transverse extension. Examples
are ferromagnetic films in saturating magnetic fields and
ferromagnetic fluids in narrow channels with a magnetic field
perpendicular to the surface. The corresponding dispersion relations
are anisotropic \cite{dam.63} and nonmonotonic \cite{bas.93},
respectively. The different types of waves in such systems are caused
by a discrete set of the transverse component $k^\perp_j$ of the wave
vector. For a wave with a lateral component $k$ the dispersion
relation $\Omega_j$ is the bulk one where $k$ is replaced by
$\sqrt{k^2+k^{\perp 2}_j}$.  Crossings between $\Omega_{j+1}$ and
$\Omega_j$ occur when $|k^\perp_{j+1}-k^\perp_j|$ is not too large,
i.e., the extension of the system in the transverse direction is not
too narrow. The strength of interaction is strongly influenced by the
kind of boundary conditions.

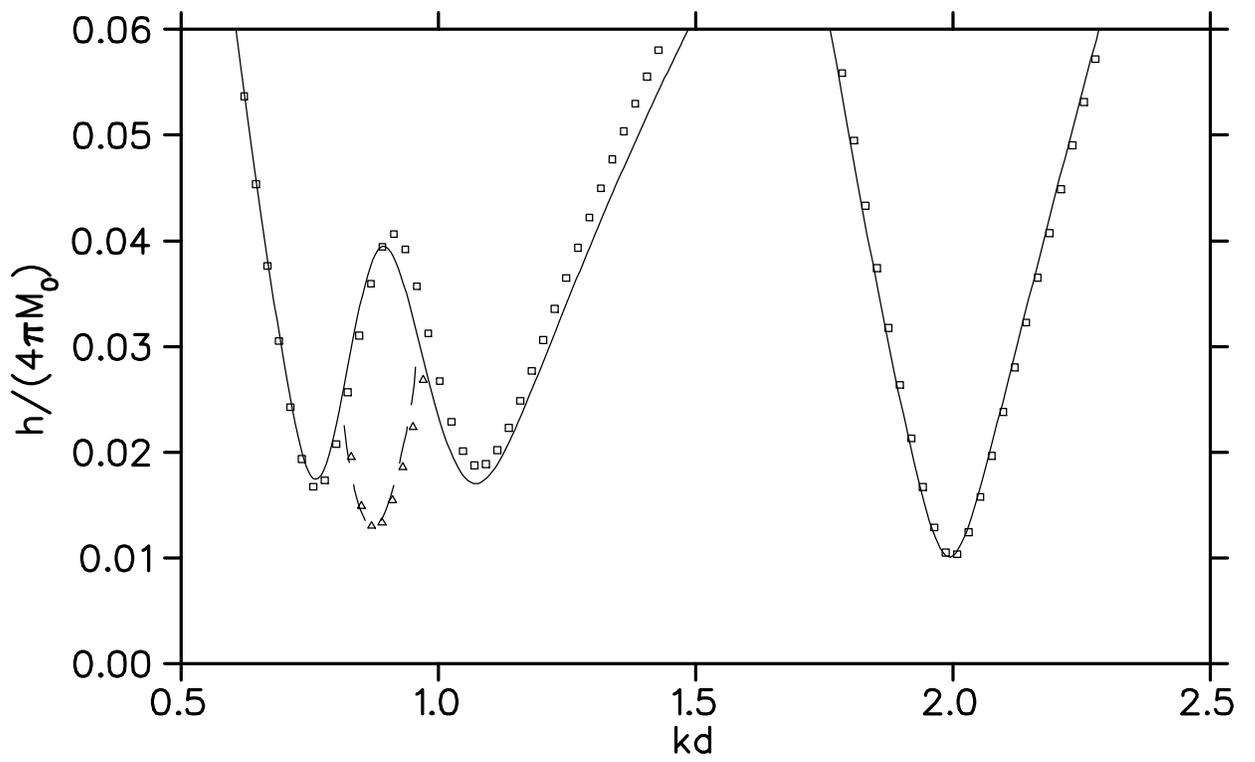
\begin{figure}
\caption[Neutral curve for FMR]{\protect\label{f.ncfmr}
The neutral curves for a parallel-pumped insulating ferromagnetic 
film of thickness $d$. The magnetic field is uniform and
perpendicular to the film plane. The values of the exchange length
$l$, the static field $H$, the Landau-Lifshitz damping $g$, and the
driving frequency $\omega$ are $0.05d$, $1.1\cdot 4\pi M_0$, $0.01$,
and $0.42\cdot 4\pi\gamma M_0$, respectively, where $M_0$ is the
magnetization and $\gamma$ the gyromagnetic ratio. The solid and
dashed lines denote the analytically found neutral curves $h_S$ and
$h_H$, respectively. The squares and triangles are numerical results
based on a Galerkin expansion \cite{elm.96}.
}
\end{figure}

We have calculated the threshold for a parallel-pumped,
ferromagnetic, insulating film in a uniform magnetic field oriented
perpendicular to the film plane \cite{rem4}. Fig.~\ref{f.ncfmr} shows
an excellent agreement between a fully numerical calculation and an
analytical one within the framework of the amplitude equations
(\ref{amp.eq}). The coefficients of the linear terms of the amplitude
equations have been already calculated in the literature
\cite{ben.74,kal.86}. On the left hand side of Fig.~\ref{f.ncfmr} we
see the neutral curves in the hybridization area of the modes $j=0$
and $j=2$. The Hopf bifurcation has clearly the lowest threshold. But
unfortunately a non-interacting wave (here the mode $j=1$ on the
right hand side) has a slightly lower threshold. Very recently
Kostylev {\em et al.\/} \cite{kos.95} have shown in a fully numerical
calculation that in tangentially magnetized films the Hopf
bifurcation can have the overall lowest threshold.

A promising physical system for experimental verification of the
direct Hopf bifurcation is the Faraday instability of a ferrofluid in
a static field perpendicular to the surface in order to get a
nonmonotonic dispersion relation. The fluid should be in an annular
container with a width which is roughly as large as, or slightly
larger, than the wavelength of the surface waves at the local maximum
of the dispersion relation. At the cross points of dispersion
relations of waves with different transverse wave numbers the group
velocities have often different signs. Thus the Hopf bifurcation
appears first if the damping is not to strong. Without any detailed
calculation we can decided whether a non-interacting wave has still a
lower threshold or not.

In this letter the influence of hybridization on the parametric
resonance of waves has been studied. We have shown that under very
general conditions the usual period-doubling bifurcation is replaced
by a direct Hopf bifurcation leading to quasi-periodic traveling
waves. The Hopf frequency is roughly given by half of the
hybridization gap. 

\acknowledgments
I gratefully acknowledge P. Talkner for his critical reading 
of the manuscript.
This work was supported by the Swiss National Science Foundation.

\begin{references}
\bibitem{rem4}
In the calculation only the exchange interaction and the dipolar
interaction of the spins are included. Furthermore unpinned
spins at the sample boundary are assumed \cite{elm.96}.
\end{references}
\end{document}